\documentclass[aipdoc,apl,twocolumn,groupedaddress,showpacs]{revtex4}
\usepackage{graphicx}
\usepackage{tabularx}
\usepackage{amsfonts}
\usepackage{amssymb}
\usepackage{amsbsy}
\begin{document}

\title{Acoustic phonon scattering in a low density, high mobility AlGaN/GaN field effect transistor}

\author{E. A. Henriksen}
\affiliation{Department of Physics, Columbia University, New York, New York 10027}
\author{S. Syed*}
\affiliation{Department of Physics, Columbia University, New York, New York 10027}
\author{Y. Ahmadian}
\affiliation{Department of Physics, Columbia University, New York, New York 10027}
\author{M. J. Manfra}
\affiliation{Bell Laboratories, Lucent Technologies, Murray Hill,
NJ 07974}
\author{K. W. Baldwin}
\affiliation{Bell Laboratories, Lucent Technologies, Murray Hill,
NJ 07974}
\author{A. M. Sergent}
\affiliation{Bell Laboratories, Lucent Technologies, Murray Hill, NJ 07974}
\author{R. J. Molnar}
\affiliation{MIT Lincoln Laboratory, Lexington, MA 02420-0122}
\author{H. L. Stormer}\affiliation{Department of Physics, Columbia University, New York, New York
10027}
\affiliation{Bell Laboratories, Lucent Technologies, Murray
Hill, NJ 07974}

\begin{abstract}
We report on the temperature dependence of the mobility, $\mu$, of the two-dimensional electron gas in a variable density AlGaN/GaN field effect transistor, with carrier densities ranging from 0.4$\times10^{12}$ cm$^{-2}$ to  3.0$\times10^{12}$ cm$^{-2}$ and a peak mobility of 80,000 cm$^{2}$/Vs.  Between 20 K and 50 K we observe a linear dependence $\mu_{ac}^{-1} = \alpha$T indicating that acoustic phonon scattering dominates the temperature dependence of the mobility, with $\alpha$ being a monotonically increasing function of decreasing 2D electron density.  This behavior is contrary to predictions of scattering in a degenerate electron gas, but consistent with calculations which account for thermal broadening and the temperature dependence of the electron screening.  Our data imply a deformation potential D = 12-15 eV. 
\end{abstract}
 \maketitle

In high quality AlGaN/GaN heterostructures the temperature dependence of the mobility of the two-dimensional (2D) electrons is determined solely by acoustic phonon scattering over a wide range of low temperatures\cite{Knap02,Hsu97}.  In this regime the inverse electron mobility, which is proportional to the scattering rates, varies linearly with temperature with the slope depending only upon the 2D electron density, and the values of the deformation potential and piezoelectric constants which characterize the strength of the electron-phonon interaction.  Therefore, by measuring the temperature dependence of the mobility across a range of densities, and using literature values of the piezoelectric constants, both the poorly known deformation potential and the density dependence of electron scattering rates due to acoustic phonons may be determined.

The density dependence of acoustic phonon scattering in GaN heterostructures has not been extensively explored.  To date, the only data on acoustic phonon scattering in the AlGaN/GaN 2D electron system (2DES) are from a small collection of fixed, moderate density samples\cite{Knap02}.  However with the recent development of high quality, high mobility AlGaN/GaN field effect transistors (FETs), it is now possible to study transport phenomena across a wide range of densities in a single device\cite{Manfra04}.  By applying a voltage to the gate of the FET, the 2D electron density, $\text{n}_{2D}$, may be continuously varied with the advantage that all other system parameters (e.g. the distribution of fixed imperfections) remain constant.  We have utilized such a high mobility FET, together with a set of fixed density heterostructures, to systematically explore the density dependence of acoustic phonon scattering rates over more than an order of magnitude in density.  This range is much broader than in previous work and so allows an effective comparison with theories of electron scattering by deformation potential and piezoelectric phonons.  We find that existing calculations of acoustic phonon scattering in degenerate GaN 2DESs are insufficient to explain our data quantitatively.  However, the main features and trends can be reproduced if the temperature dependence of electron energy and screening are taken into account.

In this study we have utilized a FET along with a set of fixed density heterostructures.  The FET device has a peak mobility of 80,000 cm$^{2}$/Vs, and a continuously tunable $\text{n}_{2D}$ that ranges over $0.4-3.0\times 10^{12}$ cm$^{-2}$, among the lowest densities measured in GaN.  The fixed density samples have $\text{n}_{2D} = 1-7\times 10^{12}$ cm$^{-2}$, and low-T mobilities of $8,000-30,000$ cm$^{2}$/Vs.  All devices were grown by molecular beam epitaxy (MBE) on GaN templates prepared by hydride vapor phase epitaxy (HVPE).  In the FET, the HVPE GaN is 40 $\mu$m thick and is grown atop a sapphire substrate.  The HVPE GaN is known to have a low density of threading dislocations of approximately $10^{8}$ cm$^{-2}$, and has been compensated with Zn at 10$^{17}$ cm$^{-3}$ to suppress any residual conductivity\cite{Manfra04}.  A layer of 1.5 $\mu$m nominally undoped MBE GaN is grown on top of the HVPE GaN, and followed by 160 \AA{} of Al$_{0.06}$Ga$_{0.94}$N.  The capping layer consists of 30 \AA{} of MBE grown GaN.  The fixed density samples were grown similarly, but with different AlGaN and GaN thicknesses in order to vary $\text{n}_{2D}$.  To make the FET, a 0.5 mm long and 80 $\mu$m wide Hall bar was patterned by photolithography, followed by a 100 nm deep etch in a Cl-based plasma.  Three voltage contacts are located on either side of the Hall bar, and an insulated gate is created on the FET surface by depositing 10 nm Ni and 100 nm Au on top of 100 nm of SiO$_{2}$.  The fixed density samples consist of 4$\times$4 mm squares with 8 contacts placed symmetrically around the edge. The contacts in all structures  consist of a Ti/Al/Ni/Au metal stack annealed at $800^{\circ}$ C for 30 s.

\begin{figure}
\includegraphics[scale=0.98]{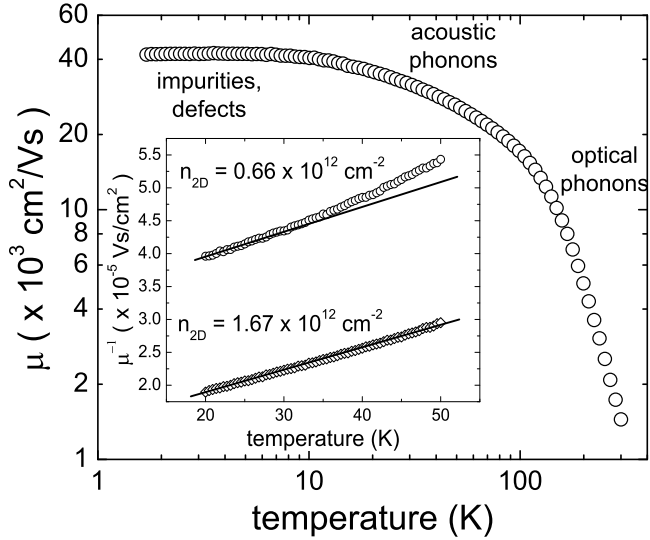}
\caption{\label{Fig.1} Typical experimental mobility vs. temperature, showing regions dominated by different types of scattering.  Inset: inverse mobilities plotted against temperature for two different densities in the FET; solid lines are a guide to the eye.}
\end{figure}

We performed low field electron density and mobility measurements in the FET and fixed density samples in the standard Hall and Van der Pauw configurations.  Fig. 1 shows a typical trace of mobility vs. temperature, in which regions limited by different scattering mechanisms are marked.  Mobility data were taken between 300 and 1.7 K in most devices.  Between 20 K and 50 K, a clear linear relationship is observed when the data are plotted as inverse mobility vs. temperature; see the inset of Fig. 1.  In some traces (e.g. the top one in the inset) the data depart from linearity as $\text{T} \to50$ K; we will return to this observation below.  In common with previous authors\cite{MPH84, Knap02} we perform a linear fit to the inverse mobility using
\begin{equation} 
\frac{1}{\mu}= \frac{1}{\mu_{0}} + \frac{1}{\mu_{ac}} =  \frac{1}{\mu_0} + \alpha T
\label{eq:one}
\end{equation}
where $\alpha$ is the slope of the fit ($\alpha = d(1/\mu) / dT$ ), $\mu_{ac}$ is the component of the mobility due to acoustic phonon scattering, and the extrapolated zero-temperature mobility $\mu_0$ accounts for all T-independent scattering mechanisms such as from remote impurities or charged dislocations.  Fig. 2 shows the slope $\alpha$ as a function of $\text{n}_{2D}$.  The significance of $\alpha$ lies in its proportionality to the acoustic phonon scattering rate, $1/\tau_{ac}$ = (e/m*) $\alpha$T.  Therefore, $\alpha$ determines the density dependence of the electron scattering rates due solely to acoustic phonons.

The upward trend of $\alpha$ at low $\text{n}_{2D}$ in Fig. 2 is striking.  It departs strongly from predictions based on a treatment of acoustic phonon scattering in a screened, degenerate 2DES (solid lines)\cite{Price84, Hsu97}. We see $\alpha$ increase monotonically by nearly a factor of two over an order of magnitude decrease in density for $\text{n}_{2D}< 2 \times 10^{12}$ cm$^{-2}$, and show no particular density dependence for $\text{n}_{2D}> 2 \times 10^{12}$ cm$^{-2}$. Data marked as $\blacktriangle$ are due to Knap \textit{et al}\cite{Knap02}, which agree with our high density data, $\text{n}_{2D} \approx 5.5 \times 10^{12}$ cm$^{-2}$.  However the lower density data of Ref[1] fall considerably below ours although they are close to the degenerate theory.  The source of this discrepancy is unknown, but is unlikely a result of the differences in scattering from fixed sources.  Such an effect is included in the term $\mu_0$, which for $\text{n}_{2D} = 2-3 \times 10^{12}$ cm$^{-2}$ is 70,000 cm$^2$/Vs in our FET and $\approx 20,000$ cm$^2$/Vs in the fixed density samples; the lower density samples of Ref[1] have $\mu_0 \approx 60,000$ cm$^2$/Vs.  

\begin{figure}
\includegraphics[scale=1]{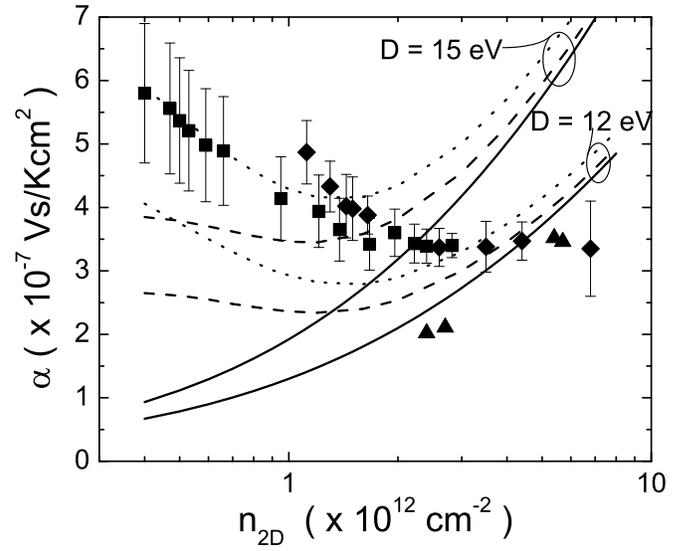}
\caption{\label{Fig.2} $\alpha$ vs. 2D electron density.  Data shown is from the FET ($\blacksquare$), fixed density samples ($\blacklozenge$), and Ref[1] ($\blacktriangle$).  Calculated fits are for two values of the deformation potential (D = 12 and 15 eV): \pmb{\textemdash} for degenerate 2DES; \pmb{\textendash{} \textendash}{} for non-degenerate 2DES, with $\text{n}_{depl}=0$; and \pmb{\textperiodcentered \textperiodcentered \textperiodcentered}{} for non-degenerate 2DES, with $\text{n}_{depl}=6 \times 10^{11}$ cm$^{-2}$.}
\end{figure}

A qualitatively similar trend in the density dependence of $\alpha$ has been previously observed in the AlGaAs/GaAs 2DES by Harris \textit{et al}\cite{Harris90}. Their data also conflicted with predictions of phonon scattering in a screened, degenerate 2DES.  This was resolved by Kawamura and Das Sarma\cite{KD92}, who noted that at lower $\text{n}_{2D}$ the Fermi temperature, which is proportional to $\text{n}_{2D}$, becomes comparable to the highest temperatures in the experiment (e.g. $\text{T}_{F} = 41$ K for $\text{n}_{2D} = 1 \times 10^{11}$ cm$^{-2}$ in GaAs; Harris \textit{et al} took data up to 40 K).  The 2DES now ceases to be degenerate and the simple linear temperature dependence of $\mu^{-1}$ is modified. To account for this, Ref[7] calculates acoustic phonon scattering in a non-degenerate 2DES by including thermal broadening of the electron distribution function and the T dependence of electron screening.  These additional T dependencies tend to decrease the mobility as $\text{T} \to \text{T}_F$, resulting in a nonlinear increase of $\mu^{-1}$ as T increases.  Therefore, a linear fit to $\mu^{-1}$  using eq. \ref{eq:one} will create a slope ($\alpha$) greater than if non-degeneracy effects were not present. This is clearly visible in the inset to Fig. 1: the top trace, for $\text{n}_{2D} = 0.66 \times 10^{12}$ cm$^{-2}$ (T$_F = 80$ K), departs from linearity toward 50 K; whereas the bottom trace, with $\text{n}_{2D} = 1.67 \times 10^{12}$ cm$^{-2}$ (T$_F = 202$ K) is linear over $20-50$ K.  

Guided by the similarity between the GaN and GaAs data, we adapted the calculations by Price\cite{Price84} and Kawamura \textit{et al} \cite{KD92} for the GaN system, using the values of the deformation potential, D, and the depletion charge density, $\text{n}_{depl}$, as adjustable parameters. All scattering is with acoustic phonons due to the deformation potential and the piezoelectric field.  In calculating the contribution from the latter, we used the form factors appropriate for scattering of 2D electrons by bulk phonons in a wurtzite crystal\cite{Ridley02}.  We used $C_l = 3.7 \times 10^{11}$ N/m$^2$ and $C_t = 1.3 \times 10^{11}$ N/m$^2$  for the longitudinal and transverse elastic constants\cite{Polian96}, with $h_{33} = 0.65$ C/m$^2$, $h_{31} = -0.37$ C/m$^2$, and $h_{15} = -0.33$ C/m$^2$ for the piezoelectric constants\cite{Shur97, Ridley02}.

In Fig. 2 we have plotted the results of our calculations for comparison with our data.  The solid lines show the theory for the degenerate case with deformation potential D = 12 and 15 eV, and zero $\text{n}_{depl}$.  The dashed and dotted lines are calculated using the non-degenerate theory of Kawamura \textit{et al} for D = 12 and 15 eV, and various $\text{n}_{depl}$.  Clearly these reflect the data much better than the degenerate theory, in particular by qualitatively capturing the rise in $\alpha$ at lower densities.  The best fits are achieved by using non-zero values for $\text{n}_{depl}$, although no single set of values for D and $\text{n}_{depl}$ reproduces $\alpha$ over the measured range of $\text{n}_{2D}$.  We achieve the best results at lower densities for $\text{n}_{depl} = 6 \times 10^{11}$ cm$^{-2}$ and D = 15 eV (upper dotted line), while for $\text{n}_{2D} > 2 \times 10^{12}$ cm$^{-2}$, the better fit is with D = 12 eV.   These values of D are considerably higher than the D = 8.3 eV calculated for bulk wurtzite GaN\cite{Chin94}, analogous to the situation in GaAs for which D = 7 eV in bulk, and 13 eV in 2D scattering\cite{KD92}.

In comparing our data with our calculations we note that piezoelectric phonon scattering becomes important at low densities\cite{Knap02}, but the piezoelectric constants $h_x$ of wurtzite GaN are not well known.  Prior to the work of Kawamura \textit{et al}, Harris \textit{et al} noted that in GaAs the rise in $\alpha$ at low densities could be reproduced by enhancing the value of $h_{14}$ (for cubic GaAs) by a factor of 1.5\cite{Harris90}.  Thus the rise of $\alpha$ in the GaN 2DES may be partly due to larger values for the $h_x$ than reported in the literature.  Also, the value of $\text{n}_{depl}$ required in the best fit seems rather high.  Assuming pinning of the Fermi level at mid-gap in the Zn-compensated HVPE GaN, the electric field across the undoped MBE GaN is equivalent to only $\text{n}_{depl} = 1.2 \times 10^{11}$ cm$^{-2}$.    Finally we note that at high $\text{n}_{2D}$, the Bloch-Gr\"{u}neisen (B-G) effect will reduce phonon scattering for $\text{T} < \text{T}_{B-G}$\cite{Ridley99}.  We have repeated our analysis on $\mu_{ac}$ vs. T for $\text{n}_{2D}=10^{13}$ cm$^{-2}$ from Fig. 1 of Ref[12], for which $\text{T}_{B-G} = 94$ K, and find that $\alpha$ is increased by $\approx 9\%$ in the B-G regime.  With $\text{T}_{B-G} \sim \text{n}_{2D}^{1/2}$, this effect rapidly diminishes with decreasing density and is not involved in the low density behavior of $\alpha$.

In conclusion, we have measured the temperature dependence of the 2D mobility over a large range of electron densities in high quality AlGaN/GaN heterostructures.  We compared these results with calculations of electron scattering by acoustic phonons, and found a qualitatively good fit only by accounting for additional temperature dependencies of the electron gas that arise when the system becomes non-degenerate at low densities.  We derive a value for the GaN acoustic deformation potential between 12 and 15 eV, with the fit improving with an \textit{ad hoc}, non-zero depletion density. It remains unclear whether the present theory, even when adjusted for more precise future piezoelectric constants, will be able to account for the experimental data. 

The authors would like to thank A. Mitra, Y. W. Tan, C. F. Hirjibeheddin, and I. Dujovne for many useful discussions.  This work is funded under ONR Project no. N00014-04-1-0028.

* Current address: Physics Dept., University of Illinois, Urbana-Champaign, IL 61801.

\end{document}